\documentclass[journal,onecolumn]{IEEEtran}
\usepackage{amsmath,amsthm}
\usepackage[T1]{fontenc}
\usepackage[tt=false,type1=true]{libertine}
\usepackage[varqu]{zi4}
\usepackage[libertine]{newtxmath}
\usepackage{booktabs}
\usepackage[nocompress]{cite}
\AtBeginDocument{}
\usepackage[hidelinks]{hyperref}

\newtheorem{theorem}{Theorem}[section]

\newtheorem{lemma}[theorem]{Lemma}
\newtheorem{proposition}[theorem]{Proposition}
\newtheorem{corollary}[theorem]{Corollary}

\newtheorem*{bconj}{Blackburn's Conjecture}
\theoremstyle{definition}
\newtheorem{definition}[theorem]{Definition}

\begin{document}

\title{A Proof of Blackburn's Conjecture\\ on Non-Overlapping Codes}

\author{Yeow Meng Chee%
\thanks{Y. M. Chee is with the Singapore University of Technology and Design, Singapore (e-mail: ymchee@sutd.edu.sg).}%
\thanks{Research supported by SUTD Grant SKI 2021\_07\_04.}}

\maketitle

\begin{abstract}
A code is non-overlapping if no nonempty proper prefix of a codeword is a suffix of a codeword. Let $S_n(q)$ be the largest size of a non-overlapping code of length $n$ over an alphabet of size $q$. Levenshtein proved that $S_n(q)\le c_nq^n$, where $c_n=(1/n)\bigl((n-1)/n\bigr)^{n-1}$. The codes $I^{n-1}J$, where $I$ and $J$ partition the alphabet, attain this bound when $n$ divides $q$, and Blackburn conjectured that for every $n$ and all sufficiently large $q$, the best such code is a maximum non-overlapping code. We prove this conjecture. The proof uses generating functions for the nested partitions that describe maximal non-overlapping codes. We also show that only the codes $I^{n-1}J$ with $|I|=(n-1)q/n$ and their reversals meet Levenshtein's bound.
\end{abstract}


\section{Introduction}\label{sec:intro}

Let $\Sigma_q$ be an alphabet of size $q$, and let $\Sigma_q^n$ be the set of words of length $n$ over $\Sigma_q$. For sets of words $A$ and $B$, we write $AB$ for the set of concatenations $ab$ with $a\in A$ and $b\in B$, and $A^k$ for the set of concatenations of $k$ words of $A$. Two words $u,v\in\Sigma_q^n$ \emph{overlap} if some nonempty proper prefix of $u$ is a suffix of $v$. A code $C\subseteq\Sigma_q^n$ is \emph{non-overlapping} if no two codewords overlap, a codeword with itself included. We write $S_n(q)$ for the largest size of a non-overlapping code in $\Sigma_q^n$. A non-overlapping code $C\subseteq\Sigma_q^n$ is \emph{maximum} if $|C|=S_n(q)$. It is \emph{maximal} if it is not properly contained in another non-overlapping code in $\Sigma_q^n$, that is, if $C\cup\{u\}$ is not non-overlapping for every $u\in\Sigma_q^n\setminus C$. Every maximum code is maximal.

Non-overlapping codes were introduced by Levenshtein~\cite{Lev64}, who showed that they are exactly the codes for which there is a decoding automaton that synchronizes regardless of its initial state. Gilbert~\cite{Gil60} studied the same class for the synchronization of binary messages. The property that no word overlaps itself, called bifix-freeness, was studied by Nielsen~\cite{Nie73}, and it is basic to the theory of frame synchronization~\cite{Mas72}. In frame synchronization a receiver searches a data stream for a marker. If several markers are used and they form a non-overlapping code, a marker cannot be detected at a position that straddles two markers, so the receiver can search for all markers simultaneously. This led to distributed sequences~\cite{dLW00} and to the systematic study of cross-bifix-free sets~\cite{BLT14,BPP12,CKPW13,BBPPS16}, which are non-overlapping codes under another name.

The same property now arises in DNA-based data storage. In the random-access architecture of Tabatabaei Yazdi et al.~\cite{YYMZM15}, each data block carries an address primer, and the block is selected by annealing the complementary primer. Primers must not overlap, so that a primer cannot anneal at a shifted position. This motivated mutually uncorrelated codes~\cite{YKGM18,LY19}, which are non-overlapping codes with further constraints such as balance and error correction. Non-overlapping codes also appear as strong comma-free codes in models of the genetic code~\cite{FMPSS23}.

In all these applications the codewords serve as distinguishable markers or addresses, so their number is the resource to be maximized, and $S_n(q)$ is the basic quantity.

\subsection{Prior Work}

Levenshtein~\cite{Lev70} proved the upper bound
\begin{equation}\label{eq:lev}
S_n(q)\le c_nq^n,\quad\text{where}\quad c_n=\frac1n\Bigl(\frac{n-1}{n}\Bigr)^{n-1}.
\end{equation}
Chee et al.~\cite{CKPW13} proved the weaker bound $S_n(q)\le q^n/(2n-1)$ and gave codes within a constant factor of optimality for each fixed $q$. Constructions based on lattice paths were given by Bilotta, Pergola and Pinzani~\cite{BPP12} and by Barcucci et al.~\cite{BBPPS16}. Blackburn~\cite{Bla15} gave a construction within a constant factor of~\eqref{eq:lev} for all $n$ and $q$. Its simplest case is the following. Partition $\Sigma_q=I\cup J$ into nonempty sets with $|I|=s$. The codewords of $I^{n-1}J$ are the words of length $n$ whose first $n-1$ symbols lie in $I$ and whose last symbol lies in $J$. Every proper prefix of a codeword has all its symbols in $I$, while every nonempty suffix ends with a symbol of $J$, so the code is non-overlapping. Its size is $s^{n-1}(q-s)$.

Let
\begin{equation}\label{eq:M}
B_n(q)=\max_{s\in\{1,2,\dots,q-1\}}s^{\,n-1}(q-s)
\end{equation}
be the largest size of such a code. We call the codes $I^{n-1}J$ \emph{Blackburn codes}. Blackburn showed that $S_n(q)=B_n(q)=c_nq^n$ whenever $n$ divides $q$, that $S_2(q)=B_2(q)$ and $S_3(q)=B_3(q)$ for all $q$, and that $S_n(q)/q^n\to c_n$ as $q\to\infty$. He conjectured the following.

\begin{bconj}[{\cite[Conjecture~1]{Bla15}}]
For every $n\ge2$ and all sufficiently large $q$, some code $I^{n-1}J$ is a maximum non-overlapping code. Equivalently, $S_n(q)=B_n(q)$ for all sufficiently large $q$.
\end{bconj}

Stanovnik, Mo\v{s}kon and Mraz~\cite{SMM24} built on an algorithm of Fimmel et al.~\cite{FMPSS23} and described all maximal non-overlapping codes by nested partitions of words. This reduced the determination of $S_n(q)$ to an integer program in the sizes of the parts. They proved $S_4(q)=B_4(q)$ for all $q$, computed $S_n(q)$ for $q\le6$, and formulated a conjecture~\cite[Conjecture~1]{SMM24} on the structure of maximum codes which would imply Blackburn's Conjecture for $q\ge n$. Related problems for variable-length codes~\cite{Bil17,WW22,WW24} and for codes in which only overlaps of certain lengths are forbidden~\cite{BEKS24} have also been studied, and for binary codes Blackburn~\cite{Bla26} recently improved Levenshtein's bound~\eqref{eq:lev} by a factor of $5/6$ as $n\to\infty$, to $\bigl(5/(6e)+O(1/n)\bigr)2^n/n$. Generating functions for words with prescribed overlaps go back to Guibas and Odlyzko~\cite{GO81}, see also Goulden and Jackson~\cite{GJ83} and Flajolet and Sedgewick~\cite{FS09}.

Blackburn's Conjecture is difficult for the following reason. When $n$ does not divide $q$, the integer optimum $B_n(q)$ of Blackburn's construction lies below Levenshtein's bound~\eqref{eq:lev} by an amount of order $q^{n-2}$ (Lemma~\ref{lem:gap}). An upper bound that proves the conjecture must therefore be sharp in its second-order term for the codes close to the best Blackburn codes, which~\eqref{eq:lev} is not.

\subsection{Results}

Our main result is the following.

\begin{theorem}\label{thm:main}
Blackburn's Conjecture is true: for every $n\ge2$ there is a $q_0(n)$ such that $S_n(q)=B_n(q)$ for all $q\ge q_0(n)$.
\end{theorem}

We prove Theorem~\ref{thm:main} in Section~\ref{sec:all}. Along the way we show that a code meets Levenshtein's bound~\eqref{eq:lev} only if, up to reversing every codeword, it is a Blackburn code with $|I|=(n-1)q/n$ (Theorem~\ref{thm:eq}). The size $s^{n-1}(q-s)$ of a Blackburn code, as a function of $s\in\mathbb R$, is largest at $s=(n-1)q/n$, where it equals $c_nq^n$. So Blackburn's Conjecture asks whether the best Blackburn code, for which $s$ must be an integer, is still at least as large as every other code. The condition that $q$ be large cannot be dropped, since the computations of Stanovnik, Mo\v{s}kon and Mraz~\cite{SMM24} for $q\le6$ give $S_n(q)>B_n(q)$ in some cases, for example $S_5(2)=2>B_5(2)=1$.

\subsection{Proof Strategy}\label{sec:strategy}

Our starting point is the description of maximal non-overlapping codes by nested partitions~\cite{FMPSS23,SMM24}. The alphabet is split into a left part and a right part, the concatenations of left words with right words are split again, and so on, and the code consists of the words of length $n$ produced in this way. If we count the left words and the right words by length with two generating functions, the whole construction reduces to a single factorization of $1-qz$ up to degree $n$, and the size of the code becomes one coefficient of a rational function. So determining $S_n(q)$ is equivalent to the problem of maximizing a coefficient, and Blackburn codes give the simplest candidates.

The proof then looks at the first split, the partition of the alphabet. If one part is large, as in Blackburn codes, we expand the rational function around its linear part. Every further term of the expansion can only decrease the size of the code, so the code is no larger than the Blackburn code with the same first split. If neither part is large, we compare the code with Levenshtein's bound instead. The generating functions give a short proof of this bound that discards a nonnegative remainder, and the remainder vanishes only for the optimal Blackburn codes. So the codes of the second kind are smaller than Levenshtein's bound by a fixed proportion, while the best Blackburn code is smaller than it only by a lower-order term, and so the best Blackburn code is larger once $q$ is large.

\section{Partition Codes and Their Generating Functions}\label{sec:gf}

This section develops the tools for the rest of the paper. 

\subsection{Notation and Terminology}\label{sec:notation}

Throughout, $n\ge2$ is the length, $q\ge2$ is the alphabet size, $c_n$ is as in~\eqref{eq:lev}, and $B_n(q)$ is as in~\eqref{eq:M}. We use the standard asymptotic notations $O$ and $\Theta$, for fixed $n$ and $q\to\infty$. For a code $C$ of length $n$ over $\Sigma_q$, we call $|C|/q^n$ the \emph{normalized size} of $C$. For a Blackburn code $I^{n-1}J$, we call $\lambda=|I|/q$ its \emph{ratio}. The Blackburn code with ratio $\lambda$ has $(\lambda q)^{n-1}(q-\lambda q)$ codewords, so its normalized size is $\beta(\lambda)$, where
\[
\beta(\lambda)=\lambda^{n-1}(1-\lambda),\quad\text{for }\lambda\in\mathbb R.
\]
Hence we have
\begin{equation}\label{eq:Mf}
B_n(q)=q^n\max_{s\in\{1,2,\dots,q-1\}}\beta\Bigl(\frac sq\Bigr).
\end{equation}
Let $\lambda_n=(n-1)/n$. Since $\beta'(\lambda)=\lambda^{n-2}\bigl((n-1)-n\lambda\bigr)$, the function $\beta$ increases on $[0,\lambda_n]$ and decreases on $[\lambda_n,1]$, and its maximum is $\beta(\lambda_n)=c_n$. So Levenshtein's bound~\eqref{eq:lev} is the maximum of $\beta$ over $\lambda\in[0,1]$. Since the ratio of a Blackburn code must be a multiple of $1/q$, the best Blackburn code is smaller than this maximum unless $\lambda_nq$ is an integer. The next lemma shows that, when $n$ does not divide $q$, the difference $c_nq^n-B_n(q)$ is of order $q^{n-2}$.

\begin{lemma}\label{lem:gap}
Let $n\ge2$. If $n$ divides $q$, then $B_n(q)=c_nq^n$. For $q$ not divisible by $n$, we have
\begin{equation}\label{eq:roundup}
c_nq^n-B_n(q)=\Theta(q^{n-2}).
\end{equation}
\end{lemma}

\begin{proof}
If $n$ divides $q$, then $q\ge n$, so the integer $s=\lambda_nq=q-q/n$ lies in $\{1,2,\dots,q-1\}$. The Blackburn code with $|I|=s$ has ratio $\lambda_n$ and normalized size $\beta(\lambda_n)=c_n$, so $B_n(q)\ge c_nq^n$, and equality holds by~\eqref{eq:lev}.

Now let $n$ not divide $q$, and let $s=\lfloor \lambda_nq\rfloor$. Since $1\le q/2\le \lambda_nq<q$, we have $s\in\{1,2,\dots,q-1\}$ and $|s/q-\lambda_n|<1/q$. Since $\beta'(\lambda_n)=0$ and $|\beta''|\le n(n-1)$ on $[0,1]$, Taylor's theorem gives $\beta(s/q)\ge c_n-\binom n2/q^2$, so~\eqref{eq:Mf} gives $c_nq^n-B_n(q)\le\binom n2q^{n-2}$.

For the lower bound, note that $\beta''(\lambda_n)=-(n-1)\lambda_n^{n-3}<0$ and $\beta(\lambda)<c_n$ for $\lambda\in[0,1]\setminus\{\lambda_n\}$. So $\bigl(c_n-\beta(\lambda)\bigr)/(\lambda-\lambda_n)^2$ extends to a continuous positive function on $[0,1]$, and it has a positive minimum $\nu_n$. For every $s\in\{1,2,\dots,q-1\}$, the integer $ns-(n-1)q=n(s-q)+q$ is nonzero, since $n$ does not divide $q$, so $|s/q-\lambda_n|\ge1/nq$ and $\beta(s/q)\le c_n-\nu_n/(nq)^2$. By~\eqref{eq:Mf}, $c_nq^n-B_n(q)\ge\nu_nq^{n-2}/n^2$, and~\eqref{eq:roundup} follows.
\end{proof}

Let $z$ be an indeterminate, and let $\mathbb R[[z]]$ be the ring of formal power series in $z$ with real coefficients. For $F\in\mathbb R[[z]]$ we write $[z^m]F$ for the coefficient of $z^m$ in $F$. A series $F$ with $[z^0]F\ne0$ is invertible in $\mathbb R[[z]]$, and $1/F$ denotes its inverse. For $F,G\in\mathbb R[[z]]$ we write $F\equiv G\pmod{z^{n+1}}$ if $[z^m]F=[z^m]G$ for $0\le m\le n$, that is, if $F$ and $G$ agree in all degrees up to $n$. This relation is preserved by sums and products, and $F\equiv G$ with $[z^0]F\ne0$ implies $1/F\equiv1/G$, since the coefficients of $1/F$ up to degree $n$ depend only on those of $F$. So congruences modulo $z^{n+1}$ can be manipulated like identities, as long as only coefficients of degree at most $n$ are read off. All manipulations of power series take place in $\mathbb R[[z]]$, so no question of convergence arises, and only polynomials are evaluated at real numbers.

\subsection{Partition Codes}\label{sec:partition}

The following construction goes back to Fimmel et al.~\cite{FMPSS23} and was analyzed by Stanovnik, Mo\v{s}kon and Mraz~\cite{SMM24}. Let $W_1=\Sigma_q$. For each $i\ge1$, partition $W_i$ into a left part $L_i$ and a right part $R_i$, and let $W_{i+1}$ be the set of concatenations of a word of some $L_j$ with a word of $R_{i+1-j}$. Thus
\begin{align}
W_i&=L_i\cup R_i,&&\text{for }i\ge1, \text{ and}\label{eq:split}\\
W_i&=\bigcup_{j=1}^{i-1}L_jR_{i-j},&&\text{for }i\ge2,\label{eq:concat}
\end{align}
where $L_jR_{i-j}=\{uv:u\in L_j,\ v\in R_{i-j}\}$. The union in~\eqref{eq:split} is disjoint by definition, and the sets $L_jR_{i-j}$ in~\eqref{eq:concat} are pairwise disjoint, since Stanovnik, Mo\v{s}kon and Mraz~\cite[Theorem~6]{SMM24} showed that every word of $W_i$ has a unique splitting point $j$. We call $(L_i,R_i)$ the \emph{$i$th level}, and for each $n\ge2$ we call $C=W_n$ the \emph{partition code} of length $n$ defined by the levels.

\begin{proposition}[{\cite[Theorem~3 and Proposition~7]{SMM24}}]\label{prop:struct}
Every partition code is non-overlapping, and every maximal non-overlapping code is a partition code.
\end{proposition}

The construction of Stanovnik, Mo\v{s}kon and Mraz is stated for $n\ge3$. For $n=2$ both statements are immediate, since a code in $\Sigma_q^2$ is non-overlapping exactly when no first symbol of a codeword is a last symbol of a codeword, so a maximal one is $L_1R_1$ for a partition $\Sigma_q=L_1\cup R_1$. Since every maximum code is maximal, it follows that $S_n(q)$ is the largest size of a partition code. Reversing every codeword maps a partition code to a partition code with its left and right parts exchanged, and preserves the non-overlapping property and the size. So we may assume without loss of generality that $|L_1|\ge|R_1|$, and we do so from now on. The Blackburn code $I^{n-1}J$ is the partition code with $L_1=I$, $R_1=J$ and $L_i=\emptyset$ for $i\ge2$.

\subsection{Generating Functions}\label{sec:gfsub}

Write $\ell_i=|L_i|$, $r_i=|R_i|$ and $w_i=|W_i|$. The code $C=W_n$ depends only on the levels up to $n$, so we discard the levels above $n$ and set $\ell_i=r_i=w_i=0$ for $i>n$. The equations that~\eqref{eq:split} and~\eqref{eq:concat} impose above level $n$ then do not hold, and the identities~\eqref{eq:fact} and~\eqref{eq:prod} that follow hold only modulo $z^{n+1}$. Moreover, since $W_n=L_n\cup R_n$, the partition of $W_n$ itself does not matter, and we adopt the convention $L_n=\emptyset$, so that $\ell_n=0$ and $r_n=w_n=|C|$. We record the left words and the right words by their generating functions by length, which are the polynomials
\[
L(z)=\sum_{m=1}^{n-1}\ell_mz^m\quad\text{and}\quad R(z)=\sum_{m=1}^{n}r_mz^m.
\]

\begin{lemma}\label{lem:fact}
For every sequence of levels, we have
\begin{align}
(1-L)(1-R)&\equiv1-qz\pmod{z^{n+1}},\quad\text{and}\label{eq:fact}\\
|C|&=[z^n]R.\label{eq:size}
\end{align}
\end{lemma}

\begin{proof}
Let $W(z)=\sum_{m=1}^nw_mz^m$ be the generating function of $\bigcup_{m\le n}W_m$ by length. The two decompositions~\eqref{eq:split} and~\eqref{eq:concat} form a recursive decomposition in the sense of Goulden and Jackson~\cite[Section~2.2]{GJ83}. By~\eqref{eq:split}, $\bigcup_{m\le n}W_m$ is the disjoint union of $\bigcup_{m\le n}L_m$ and $\bigcup_{m\le n}R_m$, so the Sum Lemma~\cite[Lemma~2.2.12]{GJ83} gives
\begin{equation}\label{eq:sum}
W=L+R.
\end{equation}
By~\eqref{eq:concat} and the disjointness of its terms, concatenation $(u,v)\mapsto uv$ is a bijection from the pairs in $\bigl(\bigcup_{m\le n}L_m\bigr)\times\bigl(\bigcup_{m\le n}R_m\bigr)$ with $|u|+|v|\le n$ onto $\bigcup_{2\le m\le n}W_m$ that adds lengths. By the Product Lemma~\cite[Lemma~2.2.14]{GJ83}, these pairs are counted by the terms of $LR$ of degree at most $n$, so we have
\begin{equation}\label{eq:prod}
W-qz\equiv LR\pmod{z^{n+1}}.
\end{equation}
Eliminating $W$ from~\eqref{eq:sum} and~\eqref{eq:prod} gives $L+R-qz\equiv LR\pmod{z^{n+1}}$, which is~\eqref{eq:fact}, and~\eqref{eq:size} holds since $r_n=|C|$.
\end{proof}

We call a pair $(L,R)$ of polynomials $L=\sum_{m=1}^{n-1}\ell_mz^m$ and $R=\sum_{m=1}^{n}r_mz^m$ with nonnegative integer coefficients a \emph{feasible pair} if $\ell_1\ge r_1>0$ and the congruence~\eqref{eq:fact} holds. By Lemma~\ref{lem:fact}, every nonempty partition code $C$ gives a feasible pair $(L,R)$ with $[z^n]R=|C|$, since $R_1=\emptyset$ would give $W_i=\emptyset$ for $i\ge2$, so that $r_1>0$. For this reason we call $[z^n]R$ the \emph{size} of a feasible pair $(L,R)$ and denote it by $|C|$, also when $(L,R)$ does not come from a code. Write $w_i=\ell_i+r_i$ for $1\le i\le n$. Comparing the coefficients of $z^i$ for $1\le i\le n$ in~\eqref{eq:fact}, we see that the congruence is equivalent to
\begin{equation}\label{eq:eqs}
w_1=q\quad\text{and}\quad w_i=\sum_{j=1}^{i-1}\ell_jr_{i-j},\quad\text{for }2\le i\le n.
\end{equation}
Since $\deg L\le n-1$, we have $|C|=r_n=w_n$. For a partition code, $w_i=|W_i|$ by~\eqref{eq:split}, and~\eqref{eq:eqs} counts the words in~\eqref{eq:concat}.

The \emph{real relaxation} is the set of pairs $(L,R)$ of polynomials $L=\sum_{m=1}^{n-1}\ell_mz^m$ and $R=\sum_{m=1}^{n}r_mz^m$ with nonnegative real coefficients such that $\ell_1\ge r_1>0$ and the congruence~\eqref{eq:fact} holds. It contains every feasible pair. As for feasible pairs, the size of an element $(L,R)$ of the real relaxation is $|C|=[z^n]R$, and we call $|C|/q^n$ its normalized size. By~\eqref{eq:eqs}, we have $w_1=q$ and $w_i=\sum_{j=1}^{i-1}\ell_j(w_{i-j}-\ell_{i-j})$ for $2\le i\le n$, so by induction on $i$ each $w_i$ is a polynomial in $q,\ell_1,\dots,\ell_{i-1}$ with integer coefficients. Hence $L$ determines $w_1,\dots,w_n$, and therefore $R$, since $r_i=w_i-\ell_i$. Thus an element $(L,R)$ of the real relaxation is determined by $L$, and we identify it with $L$. So for fixed $q$, we may take the real relaxation to be the set of polynomials $L=\sum_{m=1}^{n-1}\ell_mz^m$ with real coefficients satisfying the polynomial inequalities $0\le \ell_i\le w_i$ for $1\le i\le n-1$ and $q/2\le\ell_1<q$, which is $\ell_1\ge r_1>0$, and $|C|=w_n$ is a polynomial function on it. Scaling $\ell_i$ and $r_i$ by $q^{-i}$ maps the real relaxation for $q$ to the real relaxation for $q=1$ and divides $|C|$ by $q^n$, because~\eqref{eq:eqs} and $|C|$ are homogeneous when $\ell_i$ and $r_i$ are given degree $i$. We call $\lambda=\ell_1/q$ the \emph{ratio} of an element $L$ of the real relaxation. This agrees with the ratio of a Blackburn code, for which $\ell_1=|I|$. Since $\ell_1\ge r_1>0$ and $\ell_1+r_1=q$, we have $1/2\le\lambda<1$, and scaling preserves the ratio. We call the element with ratio $\lambda$ and $\ell_i=0$ for $2\le i\le n-1$ the \emph{real analogue} of the Blackburn code with ratio $\lambda$. It has normalized size $\beta(\lambda)$.

The congruence~\eqref{eq:fact} expresses $|C|$ through $L$ alone.

\begin{lemma}\label{lem:sol}
For every element $L$ of the real relaxation, we have
\begin{equation}\label{eq:S}
|C|=-[z^n]\,\frac{1-qz}{1-L(z)}.
\end{equation}
\end{lemma}

\begin{proof}
Since $1-L$ has constant term $1$, it is invertible, and~\eqref{eq:fact} gives $1-R\equiv(1-qz)/(1-L)\pmod{z^{n+1}}$. Taking the coefficient of $z^n$ gives $-r_n=[z^n](1-qz)/(1-L)$, and $r_n=|C|$.
\end{proof}

By Proposition~\ref{prop:struct} and Lemma~\ref{lem:sol}, $S_n(q)$ is the largest value of the coefficient~\eqref{eq:S} over the polynomials $L$ that come from partition codes, so determining $S_n(q)$ is equivalent to the problem of maximizing one coefficient of a rational function.

\subsection{Levenshtein's Bound and Its Equality Case}\label{sec:lev}

We now prove Levenshtein's bound~\eqref{eq:lev} and determine when it is attained. To prove the bound we evaluate at a real number, a root of $1-L$, and this needs an exact identity in place of the congruence~\eqref{eq:fact}. Since $L$ and $R$ are polynomials, so is $(1-L)(1-R)$, and the next lemma computes its terms of degree above $n$.

\begin{lemma}\label{lem:gf}
For every element $L$ of the real relaxation, we have
\begin{equation}\label{eq:id}
(1-L(z))(1-R(z))=1-qz+|C|z^nL(z)+E(z),
\end{equation}
where
\[
E(z)=\sum_{\substack{1\le i,j\le n-1\\ i+j>n}}\ell_ir_jz^{i+j}
\]
has nonnegative coefficients.
\end{lemma}

\begin{proof}
By~\eqref{eq:eqs}, $[z^1](L+R)=q$ and $[z^i](L+R)=w_i=[z^i]LR$ for $2\le i\le n$. Since $\deg L\le n-1$, the terms $\ell_ir_jz^{i+j}$ of $LR$ with $i+j>n$ have $i\le n-1$. Since $r_n=|C|$, those with $j=n$ sum to $|C|z^nL$, and those with $j\le n-1$ form $E$. Hence $LR=L+R-qz+|C|z^nL+E$, which is~\eqref{eq:id} after adding $1-L-R$ to both sides.
\end{proof}

\begin{theorem}\label{thm:eq}
Every element $L$ of the real relaxation satisfies $|C|\le c_nq^n$. Equality holds if and only if $L(z)=\lambda_nqz$.
\end{theorem}

\begin{proof}
Since $L(0)=0$ and $L(x)\ge \ell_1x$ for $x\in[0,\infty)$, the polynomial $1-L$ has a root $\rho\in(0,1/\ell_1]$. At $z=\rho$ the left side of~\eqref{eq:id} vanishes and $L(\rho)=1$, so since $E$ has nonnegative coefficients and $\rho>0$, we have
\begin{equation}\label{eq:rho}
|C|\rho^n=q\rho-1-E(\rho)\le q\rho-1.
\end{equation}
Dividing by $\rho^n$, we have $|C|\le(q\rho-1)\rho^{-n}=q^n\beta(1/q\rho)\le c_nq^n$, since $\beta\le c_n$ on $(0,1]$ and $\beta\le0$ on $[1,\infty)$.

Suppose equality holds. Then equality holds in~\eqref{eq:rho}, so $E(\rho)=0$. Since $\rho>0$ and every coefficient of $E$ is nonnegative, every coefficient of $E$ vanishes, so $\ell_ir_j=0$ whenever $i+j>n$. Now $w_1=q>0$, $w_2=\ell_1r_1>0$, and for $3\le i\le n-1$ the sum for $w_i$ in~\eqref{eq:eqs} contains the distinct terms $\ell_1r_{i-1}$ and $\ell_{i-1}r_1$, so since $\ell_1\ge r_1$, we have
\[
w_i\ge\ell_1r_{i-1}+\ell_{i-1}r_1\ge r_1w_{i-1}>0.
\]
Since $w_{n-1}=\ell_{n-1}+r_{n-1}>0$, either $\ell_{n-1}>0$, and then $\ell_{n-1}r_j=0$ gives $r_j=0$ for $2\le j\le n-1$, or $r_{n-1}>0$, and then $\ell_ir_{n-1}=0$ gives $\ell_i=0$ for $2\le i\le n-1$. So $L=\ell_1z$, or $r_i=0$ for $2\le i\le n-1$. In the first case $|C|=\ell_1^{n-1}(q-\ell_1)=q^n\beta(\ell_1/q)$, which equals $c_nq^n$ only for $\ell_1=\lambda_nq$. In the second case, if $L\ne\ell_1z$, then $n\ge3$, and in the same way $|C|=r_1^{n-1}(q-r_1)=q^n\beta(r_1/q)$, which equals $c_nq^n$ only for $r_1=\lambda_nq>q/2$. This contradicts $\ell_1\ge r_1$. Conversely, $L(z)=\lambda_nqz$ is the real analogue of the Blackburn code with ratio $\lambda_n$, so $|C|=q^n\beta(\lambda_n)=c_nq^n$.
\end{proof}

The proof obtains Levenshtein's bound from two inequalities, the one in~\eqref{eq:rho} and $\beta(1/q\rho)\le c_n$. The first is an equality if and only if $E(\rho)=0$, and the second if and only if $1/q\rho=\lambda_n$, that is, $\rho=n/(n-1)q$. When $n$ divides $q$, the Blackburn code $I^{n-1}J$ with $|I|=\lambda_nq$ has normalized size $\beta(\lambda_n)=c_n$, so it meets~\eqref{eq:lev}, and both inequalities are equalities for it. Indeed, here $L(z)=\lambda_nqz$, so $\rho=1/\lambda_nq=n/(n-1)q$. Theorem~\ref{thm:eq} shows that both are equalities for no other code, and for no other element of the real relaxation. In terms of codes, $L(z)=\ell_1z$ means $L_i=\emptyset$ for $2\le i\le n-1$. Then~\eqref{eq:concat} gives $W_i=L_1R_{i-1}=L_1W_{i-1}$ for $3\le i\le n$, so $C=W_n=L_1^{n-1}R_1$ is the code $I^{n-1}J$ with $I=L_1$ and $J=R_1$. A code that meets Levenshtein's bound is maximum, and hence a partition code by Proposition~\ref{prop:struct}. So Theorem~\ref{thm:eq} says that a code meets Levenshtein's bound only if, up to reversing every codeword, it is a Blackburn code with $|I|=(n-1)q/n$. In particular this requires $n\mid q$.

Levenshtein's own proof~\cite{Lev70}, see also Blackburn~\cite[Theorem~2]{Bla15}, counts the words that contain no codeword as a factor, whose generating function is the reciprocal of $1-qz+|C|z^n$, and deduces that this polynomial has a positive real root. Our proof evaluates instead at the root $\rho$ of $1-L$, and the term $E(\rho)$ that it discards is what characterizes equality.

\begin{corollary}\label{cor:relax}
The maximum of $|C|$ over the real relaxation is
\[
c_nq^n=q^n\max_{0\le \lambda\le1}\beta(\lambda),
\]
and it is attained only at the real analogue of the Blackburn code with ratio $\lambda_n$.
\end{corollary}

\begin{proof}
By Theorem~\ref{thm:eq}, $|C|\le c_nq^n$, with equality only for the stated element, and this element has $|C|=q^n\beta(\lambda_n)=c_nq^n$.
\end{proof}

\section{Proof of Blackburn's Conjecture}\label{sec:all}

This section proves Theorem~\ref{thm:main}. We assume $n\ge3$, so that $(n-2)/(n-1)\ge1/2$ in the definition of moderate codes below. 

\begin{definition}\label{def:bal}
An element of the real relaxation, or a code that gives it, is \emph{skewed} if its ratio $\lambda$ satisfies $\lambda\ge(n-2)/(n-1)$, and \emph{moderate} otherwise, that is, if $1/2\le\lambda<(n-2)/(n-1)$.
\end{definition}

\subsection{Skewed Codes}\label{sec:skew}

We first show that a skewed code is no larger than the Blackburn code with the same ratio $\lambda$, which has $q^n\beta(\lambda)$ codewords.

\begin{theorem}\label{thm:unbal}
Every skewed element $L$ of the real relaxation with ratio $\lambda$ satisfies $|C|\le q^n\beta(\lambda)$. In particular, every skewed code has at most $B_n(q)$ codewords.
\end{theorem}

\begin{proof}
Write $L(z)=\ell_1z+H(z)$, where $H$ has nonnegative coefficients and no terms of degree below $2$. Since $H(z)/(1-\ell_1z)$ has no terms of degree below $2$, its $j$th power has no terms of degree below $2j$. So each coefficient of the geometric series $\sum_{j\ge0}\bigl(H(z)/(1-\ell_1z)\bigr)^j$ is a finite sum, and this series is the inverse of $1-H(z)/(1-\ell_1z)$ in $\mathbb R[[z]]$. Hence we have
\begin{align*}
\frac{1-qz}{1-L(z)}&=\frac{1-qz}{(1-\ell_1z)\Bigl(1-\dfrac{H(z)}{1-\ell_1z}\Bigr)}\\
&=\sum_{j\ge0}H(z)^j\,\frac{1-qz}{(1-\ell_1z)^{j+1}}.
\end{align*}
The term $j=0$ contributes $[z^n](1-qz)/(1-\ell_1z)=\ell_1^n-q\ell_1^{n-1}=-\ell_1^{n-1}(q-\ell_1)$. For $j\ge1$ and $m\ge1$,
\begin{align*}
[z^m]\frac{1-qz}{(1-\ell_1z)^{j+1}}&=\binom{m+j}{j}\ell_1^m-q\binom{m+j-1}{j}\ell_1^{m-1}\\
&=\binom{m+j-1}{j}\ell_1^{m-1}\Bigl(\frac{m+j}{m}\,\ell_1-q\Bigr),
\end{align*}
and the coefficient for $m=0$ is $1$. Since $H^j$ has no terms of degree below $2j$, the coefficient $[z^n]$ of the $j$th term is a sum of products of coefficients of $H^j$ with coefficients of $(1-qz)(1-\ell_1z)^{-j-1}$ in degrees $m\le n-2j\le n-2$. For these $m$ and $j\ge1$ we have $m/(m+j)\le(n-2)/(n-1)$, so each such coefficient is nonnegative, since $\ell_1/q=\lambda\ge(n-2)/(n-1)$. As $H^j$ also has nonnegative coefficients, every term $j\ge1$ adds a nonnegative amount to $[z^n](1-qz)/(1-L)$. Hence we have $[z^n](1-qz)/(1-L)\ge-\ell_1^{n-1}(q-\ell_1)$. Since $|C|=-[z^n](1-qz)/(1-L)$ by Lemma~\ref{lem:sol}, we obtain $|C|\le \ell_1^{n-1}(q-\ell_1)=q^n\beta(\lambda)$. For a feasible pair, $\ell_1$ and $r_1=q-\ell_1$ are positive integers, so $\ell_1\in\{1,2,\dots,q-1\}$ and $\ell_1^{n-1}(q-\ell_1)\le B_n(q)$.
\end{proof}

\subsection{Moderate Codes}\label{sec:moderate}

We bound moderate codes by a compactness argument, after removing the dependence on $q$ by the scaling of Section~\ref{sec:gfsub}, which maps the real relaxation for $q$ to the real relaxation for $q=1$, divides $|C|$ by $q^n$ and preserves the ratio. For $q=1$ the ratio is $\ell_1$. Let $M_n$ be the set of elements of the real relaxation with $q=1$ and ratio at most $(n-2)/(n-1)$. It contains the moderate elements for $q=1$. It is nonempty, since it contains the real analogue of the Blackburn code with ratio $(n-2)/(n-1)$, which has $\ell_1=(n-2)/(n-1)$ and $r_1=1/(n-1)$.

\begin{lemma}\label{lem:mid}
There is a constant $\mu_n<c_n$, depending only on $n$, such that every moderate feasible pair has $|C|\le\mu_nq^n$.
\end{lemma}

\begin{proof}
We first show that $M_n$ is compact. Since $\ell_1\le(n-2)/(n-1)<1$, the condition $r_1=1-\ell_1>0$ holds automatically on $M_n$, so $M_n$ is defined by the polynomial inequalities $0\le \ell_i\le w_i$ in the coefficients of $L$, together with $1/2\le \ell_1\le(n-2)/(n-1)$, and it is closed. It is bounded, since $0\le \ell_i,r_i\le w_i$ and, since $w_1=\ell_1+r_1=1$, we have $w_i\le\sum_{j=1}^{i-1}w_jw_{i-j}$, so that $w_i$ is at most the Catalan number $\binom{2i-2}{i-1}/i$ by induction. Hence $M_n$ is compact.

Compactness implies that the polynomial $w_n$ attains its maximum $\mu_n$ on $M_n$, that is, $\mu_n=w_n$ for some element of $M_n$. By Theorem~\ref{thm:eq}, $w_n<c_n$ for every element of $M_n$, since $c_n$ requires the ratio $\lambda_n=(n-1)/n$, which exceeds $(n-2)/(n-1)$. Hence $\mu_n<c_n$. Scaling a moderate element for $q$ gives an element of $M_n$, so every moderate feasible pair has $|C|\le\mu_nq^n$.
\end{proof}

\subsection{Proof of Theorem~\ref{thm:main}}\label{sec:mainproof}

We now combine Theorem~\ref{thm:unbal} with Lemma~\ref{lem:mid}. It remains to compare $\mu_nq^n$ with $B_n(q)$, the size~\eqref{eq:M} of the best Blackburn code.

\begin{proof}[Proof of Theorem~\ref{thm:main}]
Let $n\ge3$, and let $C$ be a maximum non-overlapping code in $\Sigma_q^n$. Then $C$ is maximal, so by Proposition~\ref{prop:struct} it is a partition code, and after reversing every codeword if necessary we have $|L_1|\ge|R_1|$. Since some Blackburn code has $B_n(q)$ codewords, we have $|C|=S_n(q)\ge B_n(q)$.

By Lemma~\ref{lem:gap}, $B_n(q)/q^n\to c_n$ as $q\to\infty$. Since $\mu_n<c_n$ by Lemma~\ref{lem:mid}, we have $B_n(q)>\mu_nq^n$ for all large $q$. For such $q$, if $C$ were moderate, then Lemma~\ref{lem:mid} would give $|C|\le\mu_nq^n<B_n(q)$, a contradiction. Hence $C$ is skewed, and Theorem~\ref{thm:unbal} gives $|C|\le B_n(q)$. Therefore $S_n(q)=B_n(q)$, and the Blackburn code $I^{n-1}J$ with $|I|=s$, where $s\in\{1,2,\dots,q-1\}$ maximizes $s^{n-1}(q-s)$, is a maximum code.
\end{proof}

\section{Conclusion}\label{sec:conc}

We have proved Blackburn's Conjecture for every length: for each $n$, some Blackburn code $I^{n-1}J$ is a maximum non-overlapping code once $q$ is large. Counting partition codes with generating functions turns the determination of $S_n(q)$ into the problem of maximizing one coefficient of a rational function. Blackburn's construction is optimal whenever the first level is skewed, and the equality case of Levenshtein's bound, which also shows that Levenshtein's bound is the value of the real relaxation, separates the moderate codes from that bound. The proof gives no explicit threshold. Two related open problems remain: to determine, for each $n$, the least $q_0(n)$ in Theorem~\ref{thm:main}, and to determine all pairs $(q,n)$ with $S_n(q)>B_n(q)$.


\begin{thebibliography}{99}
\bibitem{BBPPS16} E.~Barcucci, S.~Bilotta, E.~Pergola, R.~Pinzani and J.~Succi, ``Cross-bifix-free sets generation via Motzkin paths,'' \emph{RAIRO Theor. Inform. Appl.}, vol.~50, no.~1, pp.~81--91, 2016.
\bibitem{BLT14} D.~Baji\'c and T.~Lon\v{c}ar-Turukalo, ``A simple suboptimal construction of cross-bifix-free codes,'' \emph{Cryptogr. Commun.}, vol.~6, no.~1, pp.~27--37, 2014.
\bibitem{Bil17} S.~Bilotta, ``Variable-length non-overlapping codes,'' \emph{IEEE Trans. Inf. Theory}, vol.~63, no.~10, pp.~6530--6537, 2017.
\bibitem{BPP12} S.~Bilotta, E.~Pergola and R.~Pinzani, ``A new approach to cross-bifix-free sets,'' \emph{IEEE Trans. Inf. Theory}, vol.~58, no.~6, pp.~4058--4063, 2012.
\bibitem{Bla15} S.~R. Blackburn, ``Non-overlapping codes,'' \emph{IEEE Trans. Inf. Theory}, vol.~61, no.~9, pp.~4890--4894, 2015.
\bibitem{Bla26} S.~R. Blackburn, ``A tight upper bound on non-overlapping codes,'' arXiv:2611.21913, 2026.
\bibitem{BEKS24} S.~R. Blackburn, N.~N. Esfahani, D.~L. Kreher and D.~R. Stinson, ``Constructions and bounds for codes with restricted overlaps,'' \emph{IEEE Trans. Inf. Theory}, vol.~70, no.~4, pp.~2479--2490, 2024.
\bibitem{CKPW13} Y.~M. Chee, H.~M. Kiah, P.~Purkayastha and C.~Wang, ``Cross-bifix-free codes within a constant factor of optimality,'' \emph{IEEE Trans. Inf. Theory}, vol.~59, no.~7, pp.~4668--4674, 2013.
\bibitem{dLW00} A.~J. de Lind van Wijngaarden and T.~J. Willink, ``Frame synchronization using distributed sequences,'' \emph{IEEE Trans. Commun.}, vol.~48, no.~12, pp.~2127--2138, 2000.
\bibitem{FMPSS23} E.~Fimmel, C.~J. Michel, F.~Pirot, J.-S. Sereni and L.~Str\"ungmann, ``Diletter and triletter comma-free codes over finite alphabets,'' \emph{Australas. J. Combin.}, vol.~86, no.~2, pp.~233--270, 2023.
\bibitem{FS09} P.~Flajolet and R.~Sedgewick, \emph{Analytic Combinatorics}. Cambridge: Cambridge Univ. Press, 2009.
\bibitem{Gil60} E.~N. Gilbert, ``Synchronization of binary messages,'' \emph{IRE Trans. Inf. Theory}, vol.~6, no.~4, pp.~470--477, 1960.
\bibitem{GJ83} I.~P. Goulden and D.~M. Jackson, \emph{Combinatorial Enumeration}. New York: Wiley, 1983. Reprinted by Dover, 2004.
\bibitem{GO81} L.~J. Guibas and A.~M. Odlyzko, ``String overlaps, pattern matching, and nontransitive games,'' \emph{J. Combin. Theory Ser. A}, vol.~30, no.~2, pp.~183--208, 1981.
\bibitem{Lev64} V.~I. Levenshtein, ``Decoding automata invariant with respect to the initial state,'' \emph{Problemy Kibernet.}, vol.~12, pp.~125--136, 1964 (in Russian).
\bibitem{Lev70} V.~I. Levenshtein, ``On the maximum number of words in codes without overlapping,'' \emph{Problemy Peredachi Informatsii}, vol.~6, no.~4, pp.~88--90, 1970 (in Russian).
\bibitem{LY19} M.~Levy and E.~Yaakobi, ``Mutually uncorrelated codes for DNA storage,'' \emph{IEEE Trans. Inf. Theory}, vol.~65, no.~6, pp.~3671--3691, 2019.
\bibitem{Mas72} J.~L. Massey, ``Optimum frame synchronization,'' \emph{IEEE Trans. Commun.}, vol.~20, no.~2, pp.~115--119, 1972.
\bibitem{Nie73} P.~T. Nielsen, ``A note on bifix-free sequences,'' \emph{IEEE Trans. Inf. Theory}, vol.~19, no.~5, pp.~704--706, 1973.
\bibitem{SMM24} L.~Stanovnik, M.~Mo\v{s}kon and M.~Mraz, ``In search of maximum non-overlapping codes,'' \emph{Des. Codes Cryptogr.}, vol.~92, no.~5, pp.~1299--1326, 2024.
\bibitem{WW22} G.~Wang and Q.~Wang, ``$Q$-ary non-overlapping codes: a generating function approach,'' \emph{IEEE Trans. Inf. Theory}, vol.~68, no.~8, pp.~5154--5164, 2022.
\bibitem{WW24} G.~Wang and Q.~Wang, ``On the maximum size of variable-length non-overlapping codes,'' \emph{Des. Codes Cryptogr.}, vol.~93, no.~4, pp.~871--878, 2025.
\bibitem{YKGM18} S.~M.~H. Tabatabaei Yazdi, H.~M. Kiah, R.~Gabrys and O.~Milenkovic, ``Mutually uncorrelated primers for DNA-based data storage,'' \emph{IEEE Trans. Inf. Theory}, vol.~64, no.~9, pp.~6283--6296, 2018.
\bibitem{YYMZM15} S.~M.~H. Tabatabaei Yazdi, Y.~Yuan, J.~Ma, H.~Zhao and O.~Milenkovic, ``A rewritable, random-access DNA-based storage system,'' \emph{Sci. Rep.}, vol.~5, art.~14138, 2015.
\end{thebibliography}
\end{document}